\documentclass{revtex4}
\usepackage{graphicx}
\usepackage{fancyhdr}
\usepackage{amsmath}
\usepackage{color}

\begin{document}

\title{Classically conformal $B-L$ extended Standard Model and phenomenology}

%

\author{Y. Orikasa}
\affiliation{Department of Physics, Osaka University, Toyonaka, Osaka 560-0043, JAPAN}

\begin{abstract}
Bardeen has argued that once the classically conformal invariance and 
 its minimal violation by quantum anomalies are imposed on the SM, it can 
 be free from the quadratic divergences and hence the gauge hierarchy 
 problem. 
Under the hypothesis, We investigated the minimal $B-L$ extended SM with 
 a flat Higgs potential at the Planck scale. 
In this model, the $B-L$ symmetry is radiatively broken at TeV scale. 
We studied phenomenology and detectability of the model at LHC and the 
 ILC. 
\end{abstract}

\maketitle

\thispagestyle{fancy}


\section{Introduction}
The gauge hierarchy problem is one of the most important issues 
 in the SM, which has been motivating us to 
 seek new physics beyond the SM for decades. 
The problem originates from the quadratic divergence 
 in quantum corrections to the Higgs self energy, which 
 should be canceled by the Higgs mass parameter  
 with extremely high precision when the cutoff scale is much 
 higher than the electroweak scale, say, the Planck scale. 
The most popular new physics scenario which offers 
 the solution to the gauge hierarchy problem is 
 the supersymmetric extension of the SM 
 where no quadratic divergence arises by virtue of supersymmetry.

Because of the chiral nature of the SM, the SM Lagrangian 
 at the classical level possesses the conformal invariance 
 except for the Higgs mass term closely related to 
 the gauge hierarchy problem. 
Bardeen has argued  
\cite{Bardeen} that once the classical conformal invariance 
 and its minimal violation  by quantum anomalies are imposed on the SM, 
it can be free from 
 the quadratic divergences and hence the gauge hierarchy problem.
If the mechanism  really works, we can directly 
 interpolate the electroweak scale and the Planck scale.

As was first demonstrated by Coleman and Weinberg \cite{CW} 
 for the U(1) gauge theory with a massless scalar, 
 the classically conformal invariance is broken by quantum corrections 
 in the Coleman-Weinberg (CW) effective potential and the mass scale 
 is generated through the dimensional transmutation. 
It is a very appealing feature of this scheme that associated 
 with this conformal symmetry breaking, the gauge symmetry is also 
 broken and the Higgs boson arises as a pseudo-Nambu-Goldstone boson 
 whose mass has a relationship with the gauge boson mass  
 and hence predictable (when only the gauge coupling is considered).

\section{Classically conformal B$-$L extended model}
The model we will investigate is  the minimal $B-L$ extension of the SM 
with the classical conformal symmetry \cite{IOO}. 
The $B-L$  (baryon minus lepton) number is a unique anomaly free 
global symmetry that the SM accidentally possesses and 
can be easily gauged. 
Our model is based on the gauge group 
SU(3)$_c \times$SU(2)$_L\times$U(1)$_Y\times$U(1)$_{B-L}$ 
and the particle contents are listed in Table~2. 
Here, three generations of right-handed neutrinos ($N^i$)
are necessarily introduced to make the model free from all 
the gauge and gravitational anomalies. 
The SM singlet scalar ($\Phi$) works to break the U(1)$_{B-L}$ 
gauge symmetry by its VEV, and at the same time generates 
the right-handed neutrino masses.

\begin{table}[h]
\begin{center}
\begin{tabular}{c|ccc|c}
            & SU(3)$_c$ & SU(2)$_L$ & U(1)$_Y$ & U(1)$_{B-L}$  \\
\hline
$ q_L^i $    & {\bf 3}   & {\bf 2}& $+1/6$ & $+1/3$  \\ 
$ u_R^i $    & {\bf 3} & {\bf 1}& $+2/3$ & $+1/3$  \\ 
$ d_R^i $    & {\bf 3} & {\bf 1}& $-1/3$ & $+1/3$  \\ 
\hline
$ \ell^i_L$    & {\bf 1} & {\bf 2}& $-1/2$ & $-1$  \\ 
$ N^i$   & {\bf 1} & {\bf 1}& $ 0$   & $-1$  \\ 
$ e_R^i  $   & {\bf 1} & {\bf 1}& $-1$   & $-1$  \\ 
\hline 
$ H$         & {\bf 1} & {\bf 2}& $-1/2$  &  $ 0$  \\ 
$ \Phi$      & {\bf 1} & {\bf 1}& $  0$  &  $+2$  \\ 
\end{tabular}
\end{center}
\caption{
Particle contents. 
In addition to the SM particle contents, 
the right-handed neutrino $N^i$ 
($i=1,2,3$ denotes the generation index) 
and a complex scalar $\Phi$ are introduced. 
}
\end{table}

%




The Lagrangian relevant for the seesaw mechanism is given as 
\begin{eqnarray} 
 {\cal L} \supset -Y_D^{ij} \overline{N^i} H^\dagger \ell_L^j  
- \frac{1}{2} Y_N^i \Phi \overline{N^{i c}} N^i 
+{\rm h.c.},  
\label{Yukawa}
\end{eqnarray}
where the first term gives the Dirac neutrino mass term 
after the electroweak symmetry breaking, 
while the right-handed neutrino Majorana mass term 
is generated through the second term associated with 
the $B-L$ gauge symmetry breaking. 
Without loss of generality, we here work on the basis
where the second term is diagonalized and 
$Y_N^i$ is real and positive.

Under the hypothesis of the classical conformal invariance of the model, 
the classical scalar potential is described as

We assume the classically conformal invariance.
The classically conformal invariance protect the mass term of Higgs boson.
\begin{equation}
  V(\Phi, H) = \lambda_H (H^\dagger H)^2 + \lambda (\Phi^\dagger \Phi)^2 
   + \lambda^\prime (\Phi^\dagger \Phi) (H^\dagger H).  
\label{potential}
\end{equation}
We assume the following conditions at the Planck scale. \cite{IO} 
\begin{eqnarray}
 \lambda_H=\lambda^\prime=0\\
 g_{B-L}\sim g_Y. 
\end{eqnarray}
And the gauge mixing vanishes at EW scale. 
We call the Eq.(3) a flat Higgs potential. 
The Higgs quartic coupling ($\lambda_H$) and the mixing ($\lambda^\prime$) 
 are generated by quantum effect. 
We get very small and negative 
 $\lambda^\prime$($\lambda^\prime(m_{EW})\sim-{\cal O}(10^{-3})$).

Therefore there is no symmetry breaking at the classical 
 level. 
We need the Coleman-Weinberg(CW) Mechanism. 
It is a radiative symmetry breaking mechanism. 
We can consider the SM part and $B-L$ part separately, because the mixing 
 is very small. 
First we consider the $B-L$ sector in this model.
$B-L$ symmetry is broken by CW mechanism.

Once the $B-L$ symmetry is broken, the SM Higgs doublet mass 
is generated through the mixing term between $H$ and $\Phi$ 
in the scalar potential, 
\begin{equation}
  m_h^2 = -\lambda^\prime M^2.  
\label{Hmass}
\end{equation}
Where M is VEV of $\Phi$. 
The electroweak symmetry is broken in the same way as in the SM. 

The scale of $B-L$ symmetry breaking is written as a function of 
 $\lambda^\prime$ and $m_h$, 
\begin{equation}
M=\sqrt{\frac{m_h^2}{-\lambda^\prime}}. 
\end{equation}
According to our assumption, $\lambda^\prime$ is around 
 ${\cal O}(10^{-3})$. 
Therefore the $B-L$ breaking scale is around a few TeV.

\section{Phenomenology of TeV Scale $B-L$ Model}

Based on the simple assumption of a flat Higgs potential, 
we have proposed a minimal phenomenologically 
viable model with an extra gauge symmetry. 
The naturalness of the SM Higgs boson mass constrains
the $B-L$ breaking scale to be around TeV and hence, 
the mass scale of new particles in the model, 
$Z^\prime$ boson, right-handed Majorana neutrinos and 
the SM singlet Higgs boson, is around TeV or smaller. 
These new particles may be discovered at future collider 
experiments such as the LHC and ILC. 
Now we study phenomenology of these new particles.

\begin{figure}[h]
\centering
\includegraphics[width=80mm]{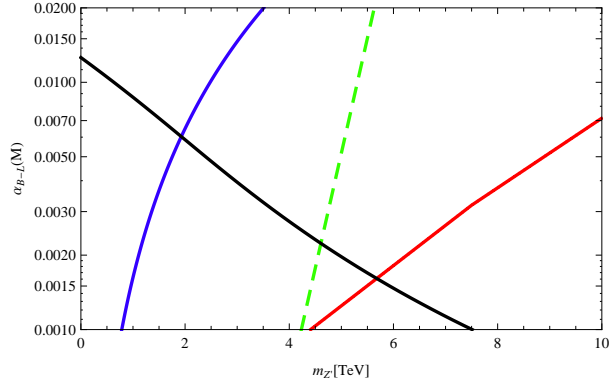}
\caption{
Model prediction is drawn in the black line (from top left to down right).
The $B-L$ gauge coupling $\alpha_{B-L}$ and  the  gauge boson mass $m_{Z'}$
are related because of the {\it flat potential assumption at the Planck scale}. 
The left side of the most left solid line in blue has been already excluded by 
 the LEP experiment.
The left of the dashed line can be explored in the 5-$\sigma$ significance 
 at the LHC with $\sqrt{s}$=14 TeV and an integrated 
 luminosity 100 fb$^{-1}$.
The left of the most right solid line (in red) can be explored at the ILC with 
 $\sqrt{s}$=1 TeV, assuming 1\% accuracy. 
} \label{FigRegion}
\end{figure}


The Fig.\ref{FigRegion} shows the prediction of our model and search regions at colliders. 
The black line is the prediction of our model. 
We have only one parameter that are important in the dynamics of the model. 
$Z^\prime$ boson mass can be written as a function of $B-L$ gauge coupling. 
The blue line is the LEP bound. 
The left side of the blue line has been already excluded by the LEP2 experiment. 



We first investigate the $Z^\prime$ boson production at the LHC. 
In our study, we calculate the dilepton production cross sections 
through the $Z^\prime$ boson exchange together with the SM processes 
mediated by the $Z$ boson and photon.
The dependence of the cross section on the final state 
dilepton invariant mass $M_{ll}$ is described as
\begin{eqnarray}
 \frac{d \sigma (pp \to \ell^+ \ell^- X) }
 {d M_{ll}}
 &=&  \sum_{a, b}
 \int^1_{-1} d \cos \theta
 \int^1_ \frac{M_{ll}^2}{E_{\rm CMS}^2} dx_1
 \frac{2 M_{ll}}{x_1 E_{\rm CMS}^2}   \nonumber \\
&\times & 
 f_a(x_1, Q^2)
  f_b \left( \frac{M_{ll}^2}{x_1 E_{\rm CMS}^2}, Q^2
 \right)  \frac{d \sigma(\bar{q} q \to \ell^+ \ell^-) }
 {d \cos \theta},
\label{CrossLHC}
\end{eqnarray}
where $E_{\rm CMS} =14$ TeV is the center-of-mass energy of the LHC.
In our numerical analysis, we employ CTEQ5M~\cite{CTEQ} for the parton 
distribution functions with the factorization scale $Q= m_{Z^\prime}$.

\begin{figure}[ht]\begin{center}
{\includegraphics*[width=.6\linewidth]{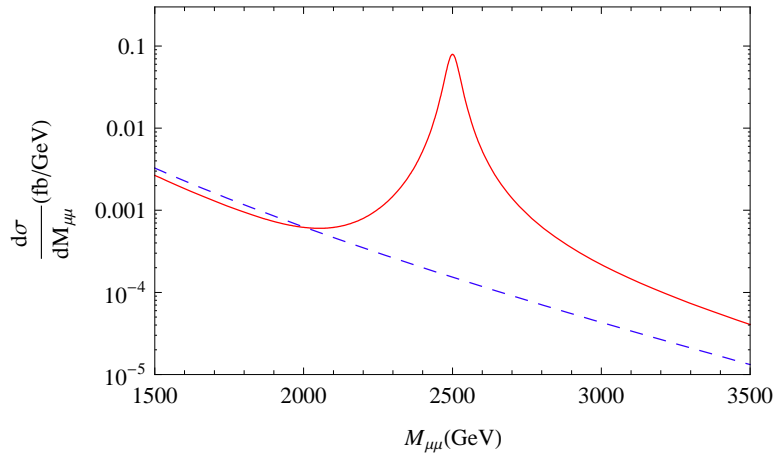}}
\caption{
The differential cross section for $pp \to \mu^+ \mu^- X$ 
at the LHC for $m_{Z^\prime}=2.5$ TeV and $\alpha_{B-L}=0.008$. 
}
\label{FigLHC}
\end{center}
\end{figure}

Fig.~\ref{FigLHC} shows the differential cross section for $pp \to
\mu^+\mu^-$ for $m_{Z^\prime}=2.5$ TeV together with the SM cross 
section mediated by the $Z$-boson and photon. 
Here, we have used $\alpha_{B-L}=0.008$ and all three right-handed 
Majorana neutrino masses have been fixed to be 200 GeV as an example. 
The result shows a clear peak of the $Z^\prime$ resonance. 
When we choose a kinematical region for the invariant mass in the range,
$M_{Z^\prime}- 2 \Gamma_{Z^\prime} \leq M_{ll} \leq 
M_{Z^\prime} + 2 \Gamma_{Z^\prime}$ with 
$\Gamma_{Z^\prime} \simeq 53$ GeV, for example, 
$560$ signal events would be observed with an integrated luminosity 
of 100 fb$^{-1}$, while only a few evens are expected 
for the SM background. 
We can conclude that the discovery of the $Z^\prime$ boson with 
mass around a few TeV and the $B-L$ gauge coupling comparable 
to the SM gauge couplings is promising at the LHC.

In order to evaluate the search reach of the $Z^\prime$ boson 
at the LHC, more elaborate study is necessary. 
We refer recent studies in \cite{Bassoetal}. 
In Fig.~\ref{FigRegion}, the dashed line (in green) shows the 5-$\sigma$ 
discovery limit obtained in \cite{Bassoetal} for $E_{\rm CMS} =14$ TeV 
with an integrated luminosity of 100 fb$^{-1}$. 
If the $B-L$ gauge coupling is comparable to the SM ones, 
$\alpha_{B-L}={\cal O}(0.01)$, the LHC can cover the region 
$m_{Z^\prime} \lesssim 5$ TeV.

Once a resonance of the $Z^\prime$ boson has been discovered 
at the LHC, the $Z^\prime$ boson mass can be determined from 
the peak energy of the dilepton invariant mass. 
After the mass measurement, we need more precise measurement 
of the $Z^\prime$ boson properties such as couplings with 
each (chiral) SM fermion, spin and etc., in order to discriminate 
different models which predict electric-charge neutral gauge bosons. 
It is interesting to note that the ILC is capable for this task 
even if its center-of-mass energy is far below the $Z^\prime$ 
boson mass~\cite{ILC}. 
In fact, the search reach of the ILC can be beyond the LHC one.

\begin{figure}[ht]\begin{center}
{\includegraphics*[width=.6\linewidth]{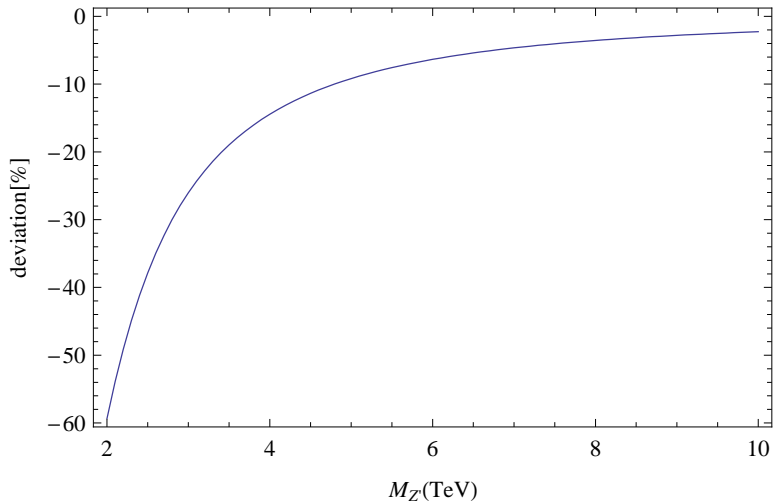}}
\caption{
Deviation (in units of \%) from the SM cross section 
as a function of $M^\prime$, 
for $\alpha_{B-L}=0.01$. 
}
\label{FigILC}
\end{center}
\end{figure}

We calculate the cross sections of the process $e^+ e^- 
\to \mu^+ \mu^-$ at the ILC with a collider energy 
$\sqrt{s}=1$ TeV for various $Z^\prime$ boson mass. 
The deviation of the cross section in our model from the SM one, 
\begin{eqnarray}
 \frac{\sigma(e^+ e^- \to \gamma, Z, Z^\prime \to \mu^+ \mu^-)}
{\sigma_{SM}(e^+ e^- \to \gamma, Z \to \mu^+ \mu^-)}-1,  
\end{eqnarray}
is depicted in Fig.~\ref{FigILC} as a function of $m_{Z^\prime}$. 
Here we have fixed $\alpha_{B-L}=0.01$ and the differential cross 
section is integrated over a scattering angle 
$-0.95 \leq \cos \theta \leq 0.95$. 
Even for a large $Z^\prime$ boson mass, for example, 
$m_{Z^\prime}=10$ TeV, Fig.~\ref{FigILC} shows a few percent deviations, 
which is large enough for the ILC with an integrated luminosity 
500 fb$^{-1}$ to identify. 
Assuming the ILC is accessible to 1 \% deviation, 
the search limit at the ILC has been investigated in \cite{Bassoetal} 
and in Fig.~\ref{FigRegion}, the red line shows the result. 
The ILC search limit is beyond the one at the LHC. 

\bigskip 

\end{document}